%% file: EarlyMig_astroph.tex
\documentclass[traditabstract]{aa}

\usepackage{graphicx}
\usepackage{txfonts}
\usepackage{amssymb}

\newcommand{\hide}[1]{} % use to omit source lines

\begin{document}

\title{The Effect of an Early Planetesimal-Driven Migration of the Giant Planets on Terrestrial Planet Formation}

\author{Kevin J. Walsh
\inst{1}
\and Alessandro Morbidelli
\inst{1}
}

\institute{University Nice Sophia Antipolis, Laboratoire Cassiop\'ee, Observatoire de la C\^ote d'Azur
BP 4229 06304 Nice cedex 04 - France -}
%TBD

\abstract
{
The migration of the giant planets due to the scattering of
planetesimals causes powerful resonances to move through the asteroid
belt and the terrestrial planet region. Exactly when and how the giant
planets migrated is not well known. In this paper we present results
of an investigation of the formation of the terrestrial planets during
and after the migration of the giant planets. The latter is assumed to
have occurred immediately after the dissipation of the nebular disk --
i.e. ``early'' with respect to the timing of the Late Heavy
Bombardment (LHB). The presumed cause of our modeled early migration
of the giant planets is angular mometum transfer between the planets
and scattered planetesimals.

Our model forms the terrestrial planets from a disk of material which
stretchs from 0.3-4.0 AU, evenly split in mass between planetesimals
and planetary embryos. Jupiter and Saturn are initially at 5.4 and 8.7
AU respectively, on orbits with eccentricities comparable to the
current ones, and migrate to 5.2 and 9.4 AU with an $e$-folding time
of 5~Myr.

Unfortunately, the terrestrial planets formed in the simulations are
not good analogs for the current solar system, with Mars typically
being much too massive. Moreover, the final distribution of the
planetesimals remaining in the asteroid belt is inconsistent with the
observed distribution of asteroids. This argues that, even if giant
planet migration had occurred early, the real evolution of the
giant planets would have to have been of the ``jumping-Jupiter''
type, i.e. the increase in orbital separation between Jupiter and
Saturn had to be dominated by encounters between Jupiter and a third,
Neptune-mass planet.  This result was already demonstrated for late
migrations occuring at the LHB time by Brasser et al. (2009) and
Morbidelli et al. (2010), and this paper shows their conclusions hold
for early migration as well.  }

\titlerunning{Effects of Early Giant Planet Migration}

\maketitle

%%% ----------------------------------------------------------------------

\section{Introduction}

The formation of the terrestrial planets is expected to have occurred
from a disk of planetesimals in two steps. In the first step, Moon to
Mars-size ``planetary embryos'' formed by runaway and oligarchic
accretion (Greenberg et al. 1978; Wetherill \& Stewart 1993; Kokubo \&
Ida 1998). In the second step, the terrestrial planets formed by
high-velocity collisions among the planetary embryos (Chambers \&
Wetherill 1998; Agnor et al. 1999; Chambers 2001; Raymond et al.
2004, 2005, 2006, 2007; O'Brien et al. 2006; Kenyon \& Bromley 2006).

The most comprehensive effort to date in modeling terrestrial planet
formation (Raymond et al. 2009) focused on 5 constraints of the
terrestrial planets: 1. the orbits, particularly the small
eccentricities, 2. the masses, with the small mass of Mars the most
difficult to match, 3. formation timescales, 4. bulk structure of the
asteroid belt and 5. the water content of Earth. Despite success with
some of these constraints in each simulation, no simulation satisfied
all the constraints simultaneously. For the simulations with fully
formed Jupiter and Saturn on nearly circular orbits, the constraint
consistently missed is the small mass of Mars. A Mars of the correct
size is only obtained in simulations where the giant planets are on
orbits with current semimajor axes but much larger
eccentricities. This scenario, however, raises the problem of not
allowing any water delivery to Earth from material in the outer
asteroid belt region. The size of Mars has been a consistent problem
for previous works with giant planets assumed on current orbits and
disks of planetesimals and embryos stretching from $\sim$0.5--4.0 AU
(Chambers \& Wetherill 1998; O'Brien et al. 2006), or even only up to
1.5 or 2.0 AU (Kokubo et al. 2006; Chambers 2001).

However, Hansen (2009) had great success creating analogs of Mars in
simulations which begin with a narrow annulus of planetary embryos
between 0.7 and 1.0 AU. In these simulations both Mercury and Mars are
formed from material that is scattered out of the original annulus by
the growing Earth and Venus analogs. In addition, the orbits of the
Earth and Venus analogs have eccentricities and inclinations similar
to those observed today and the accretion timescales are in agreement,
although on the low side, with the ages of the Earth-Moon system
deduced from the $^{182}$Hf - $^{182}$W chronometer. This model points
to the need for a truncated planetesimal disk at, or near, the
beginning of the process of terrestrial planet formation. The origin
of this truncation remains to be understood. Similarly, it remains to
be clarified how the truncation of the disk of planetesimals at 1 AU
can be compatible with the existence of asteroids in the 2-4 AU
region.

Nagasawa et al. (2005) and Thommes et al. (2008) effectively
produced a cut in the planetesimal distribution at ~1.5 AU by assuming
that the giant planets were originally on their current orbits and
that secular resonances swept through the asteroid belt during
gas-dissipation. However, the assumption that the giant planets orbits
had their current semimajor axes when the gas was still present is no
longer supported. When embedded in a gas disk, planets migrate
relative to each other until a resonance configuration is achieved
(Peale \& Lee 2002; Kley et al. 2009; Ferraz-Mello et al. 2003;
Masset \& Snellgrove 2001; Morbidelli \& Crida 2007; Pierens \&
Nelson 2008). Thus it is believed that the giant planets were in
resonance with each other when the gas disk disappeared (Morbidelli et
al. 2007; Thommes et al. 2008; Batygin \& Brown 2010) which causes
problems in understanding the consequences of the Thommes et al.
(2008) model. Moreover, the Nagasawa et al. (2005) and Thommes et
al. (2008) simulations produce the terrestrial planets too quickly
($\sim 10$~Myr), compared to the timing of moon formation indicated by
the $^{182}$Hf - $^{182}$W chronometer ($>30$~Myr and most likely
$>50$~Myr; Kleine et al. 2009) and they completely deplete the
asteroid belt by the combination of resonance sweeping and gas-drag
(see also Morishima et al. 2010, for a discussion).

The resonant configuration of the planets in a gas disk is extremely
different from the orbital configuration observed today.
Planetesimal-driven migration is believed to be the mechanism by which
the giant planets acquired their current orbits after the gas-disk
dissipation.  In fact, work by Fernandez \& Ip (1984) found that
Uranus and Neptune have to migrate outward through the exchange of
angular momentum with planetesimals that, largely, they scatter
inward. Similarly, Saturn suffers the same fate of outward migration,
though Jupiter migrates inward as it ejects the planetesimals from the
solar system. The timescale for planetesimal-driven migration of
  the giant planets depends on the distribution of the planetesimals
  in the planet-crossing region. It is typically 10~My, with 5~My as
  the lower bound (Morbidelli et al. 2010). Close encounters between
pairs of giant planets might also have contributed in increasing the
orbital separations among the giant planets themselves (Thommes et
al. 1999; Tsiganis et al. 2005; Morbidelli et al. 2007; Brasser et
al. 2009; Batygin \& Brown 2010).  Beyond the consequences for the
scattered planetesimals, the migration of the giant planets affects
the evolution of the solar system on a much larger scale, through the
sweeping of planetary resonances through the asteroid belt region.

The chronology of giant planet migration is important for the general
evolution of the solar system, including the formation of the
terrestrial planets. It has been recently proposed (Levison et al.
2001; Gomes et al. 2005; Strom et al. 2005) that the migration of the
giant planets is directly linked in time with the so-called ``Late
Heavy Bombardment'' (LHB) of the terrestrial planets (Tera et al.
1974; Ryder 2000, 2002; Kring \& Cohen 2002).  If this is true, then
the migration of the giant planets should have occurred well after the
formation of the terrestrial planets.  In fact, the radioactive
chronometers show that the terrestrial planets were completely formed
100~Myr after the condensation of the oldest solids of the solar
system (the so-called calcium alluminum inclusions, which solidified
4.568~Gyr ago; Bouvier et al. 2007; Burkhardt et al. 2008), whereas the
LHB occurred 3.9--3.8~Gyr ago. Thus the terrestrial planets should
have formed when the giant planets were still on their pre-LHB orbits:
resonant and quasi-circular. However, the simulations of Raymond et
al. (2009) fail to produce good terrestrial planet analogs when using
these pre-LHB orbits.

The alternative possibility is that giant planet-migration occurred as
soon as the gas-disk disappeared. In this case, it cannot be a cause
of the LHB (and an alternative explanation for the LHB needs to be
found; see for instance Chambers 2007). However, in this case giant
planet migration would occur while the terrestrial planets are
forming, and this could change the outcome of the terrestrial planet
formation process. In particular, it is well known that, as Jupiter
and Saturn migrate, the strong $\nu_6$ secular resonance sweeps
through the asteroid belt down to $\sim 2$~AU (Gomes 1997).  The
$\nu_6$ resonance occurs when the precession rate of the longitude of
perihelion of the orbit of an asteroid is equal to the mean precession
rate of the longitude of perihelion of Saturn, and it affects the
asteroids' eccentrcities. If the giant planet migration occurs on a
timescale of 5--10~Myr, typical of planetesimal-driven migration, then
the $\nu_6$ resonance severely depletes the asteroid belt region
(Levison et al. 2001; Morbidelli et al. 2010). This can effectively
truncate the disk of planetesimals and planetary embryos, leaving it
with an outer edge at about 1.5~AU.  Although the location of this
edge is not as close to the sun as assumed in Hansen (2009) (1~AU), it
might nevertheless help in forming a Mars analog,  i.e.
signficantly less massive than the Earth.

An equally important constraint is the resulting orbital distribution
of planetesimals in the asteroid belt region, between 2--4 AU. After
that region has been depleted of planetesimals and embryos by the
sweeping resonances, what remains will survive without major
alteration and should compare favorably with todays large
asteroids. Studies of late giant planet migration start with an
excited asteroid belt, where inclinations already vary from
0--20$^{\circ}$ (Morbidelli et al. 2010), and cannot match the
inclination distribution of the inner asteroid belt with 5~Myr or
longer migration timescales. The early migration presented here is
different because it occurs immediately after the dissipation of the
gas disk so that the planetesimal orbits are dynamically cold, with
inclincations less than 1$^{\circ}$. Thus, in principle, an early
giant planet migration could lead to a different result. Also, the
embryos will be present, another difference with late migration
scenarios.

The purpose of this paper is to investigate, for the first time,
the effect that an {\it early} migration of the giant planets could
have had on the formation of the terrestrial planets and on the final
structure of the asteroid belt. In Section~2 we discuss our methods
and in Section~3 we present our results. The conclusions and a
discussion on the current state of our understanding of terrestrial planet
formation will follow in Section~4.

\section{Methods}
We assume in our simulations that the nebular gas has dissipated,
Jupiter and Saturn have fully formed; in the terrestrial planet and
asteroid belt region, in the range 0.5-4.0 AU, the planetesimal disk
has already formed planetary embyros accounting for half of its total
mass. The lifetime of the circumstellar gas disk is observed to
  be 3--6 Myr, and both Jupiter and Saturn are expected to be fully
  formed by this time (Haisch et al. 2001). The timescales for
  oligarchic growth is similar, with lunar to Mars sized embryos
  growing on million year timescales (Kokubo \& Ida 1998,2000).

The numerical simulations are done using SyMBA, a symplectic $N$-body
integrator modified to handle close encounters (Duncan et al.
1998). In our model, the planetary embryos interact with each other;
the planetesimals interact with the embryos but not with themselves;
all particles interact with the giant planets and, except when
specified (explained further below), the giant planets feel the
gravity of embryos and planetesimals.  Collisions between two bodies
result in a merger conserving linear momentum. It has been
demonstrated by Kokubo \& Genda (2010) that this a priori assumption
of simple accretion does not significantly affect the results. The
SyMBA code has already been used extensively in terrestrial planet
formation simulations (Agnor et al.  1999; Levison \& Agnor 2003;
O'Brien et al. 2006; McNeil et al.  2005).

\subsection{Protoplanetary disk}
The initial protoplanetary disks are taken directly from O'Brien et
al. (2006), which themselves were based on those of Chambers
(2001). The O'Brien et al. (2006) study produced some of the best
matches for terrestrial planets and by using similar intial conditions
allows a direct comparison. The initial conditions are based on a
``minimum mass'' solar nebula, with a steep surface denstiy profile.
The solid mass is shared between many small planetesimals and a
  small number of large bodies, the embryos, as suggested by
  runaway/oligarchic growth simulations (Kokubo \& Ida 1998, Kokubo \&
  Ida 2000). In theory, it is possible that, by the time the gas
  disappears from the disk (which cooresponds to time zero in our
  simulations) the planetary embryos in the terrestrial planet region
  could have grown larger than the mass of Mars. However, the current
  mass of Mars seemingly excludes this possibility, and argues for
  masses to have been martian or sub-martian in mass.

The surface density profile is $\Sigma(r) = \Sigma_{0}(\frac{r}{1
  \mathrm{AU}})^{-3/2}$, where $\Sigma_{0}$ = 8 g cm$^{-2}$. The
distribution of material drops linearly between 0.7 and 0.3 AU. Half
of the mass is in the large bodies, of which there were either 25
embryos, each of 0.0933 Earth masses ($M_\oplus$) or 50 embryos of
0.0467 $M_\oplus$. The small bodies are 1/40 as massive as the large
embyros, or 1/20 as massive as the small embryos. For all test
  cases the embryos are spaced between 4--10 mutual hill radii at the
  beginning of the simulations. In some tests, the smaller
planetesimals with an initial semimajor axis larger than 2.0
were cloned into two particles with identical semi-major axis, half
the mass in each, and different random eccentricities and inclinations
(noted as 'Double Asteroids' in Table 1.). The initial eccentricities
and inclinations were selected randomly in the range of 0-0.01 and
0-0.5 degrees respectively. Thus the initial mass of the disk
consisted of 2.6 $M_\oplus$ located inside of 2~AU and a total mass of
4.7 $M_\oplus$.

\subsection{Giant planets and migration}
In all tests Jupiter and Saturn were started on orbits closer to each
other than at the present time, i.e. with semimajor axes of 5.4 and
8.7 AU, respectively. These initial orbits are just beyond their
mutual 1:2 mean motion resonance, i.e. the corrseponding ratio of
orbital periods of Saturn and Jupiter is slightly larger than 2. Even
if the giant planets should have started from a resonant configuration
- probably the 2:3 resonance (Masset \& Snellgrove 2001; Pierens \&
Nelson 2008) - it is known that secular resonance sweeping through the
asteroid belt is important only when the planets' orbital period ratio
is larger than 2 (Gomes 1997; Brasser et al. 2009). Thus, our choice
of the initial orbits of Jupiter and Saturn is appropriate for the
purposes of this study.

Each planet was forced to migrate by imposing a change to their
orbital velocities that evolves with time $t$ as:
$$v(t) = v_0 + \Delta v [1 - \exp(-t/\tau)]$$
appropriate $\Delta v$ to achieve the required change in semimajor
axis, and $\tau=5$~Myr. The latter is the minimum timescale at which
planetesimal-driven migration can occur, simply due to the lifetime of
planetesimals in the giant planet crossing region, as discussed
extensively in Morbidelli et al. (2010).  Longer timescales are
possible, but previous work has shown that fast timescales affect the
asteroid belt region less, and since terrestrial planet formation
timescales are in the 10's of millions of years, more rapid migration
has a greater chance of affecting the accretion of Mars.  Thus, we
think that restricting ourselves to the 5~My timescale is sufficient,
as this timescale is the most favorable for these purposes.

\begin{figure}
\includegraphics*[width=9.0cm,angle=0]{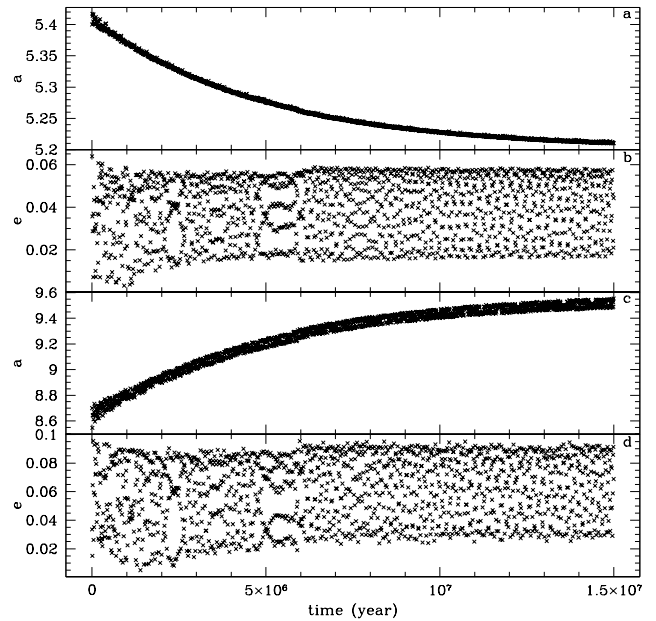}
\caption{Example of idealized migration for a system with only
  Jupiter and Saturn, ending with orbits very close to the current ones.
  Panel (a) shows the semimajor axis of Jupiter, (b) the eccentricity
  of Jupiter, (c) the semimajor axis of Saturn and (d) the eccentricity
  of Saturn, all plotted as a function of time in years.}
\label{migration}
\end{figure}

If the motion of the giant planets was not affected by the other
  bodies in the system, the evolution of the eccentricities and
  inclinations would not change much during migration (Brasser et al.,
  2009, and Fig. 1). Thus it is relatively simple to find initial
  conditions that lead to the final orbital configurations with
  eccentricities and inclinations with mean values and amplitude of
  oscillations similar to current one. In fact, as shown in Brasser et
  al. (2009), the initial values $(e_J,e_S)=(0.012,0.035)$ and
  $(i_J,i_S)=(0.23^\circ,1.19^\circ)$, after migration, lead to
  eccentricities and inclinations whose mean values and amplitudes of
  oscillation closely resemble those characterizing the current
  secular dynamics of the giant planets (see Fig.~\ref{migration}).

In our case, however, as the giant planets migrate, they scatter
planetesimals and planetary embryos, and their orbits are affected in
response. Thus, the final orbits are not exactly like those of
Fig.~\ref{migration}. Typically, for instance, the eccentricities and
inclinations of the planets are damped, and their relative migration
is slightly more pronounced than it was intended to be. Thus, we tried
to modify the initial eccentricities of Jupiter and Saturn and the
values of $\Delta v$ in order to achieve final orbits as similar as
possible to those of Fig.~\ref{migration}. However, while the effect
of planetesimals on the planets is statistically the same from
simulation to simulation, (and so can be accounted for by modifying
the initial conditions of the planets), the effects of embryos are
dominated by single stochastic events. Thus, it is not possible to
find planetary initial conditions that lead systematically to good
final orbits. In some cases the final orbits are reasonably close to
those of the current system, but in many cases they are not. In total
we performed 30 simulations. We discarded the simulations with
unsuccessful final orbits, and kept only those (9/30) that lead to
orbits resembling the current ones. These successful runs are called
hereafter ``normal migration simulations''. Our criterion for
discriminating good from bad final orbits was determined after the
15~Myr of migration, and the semimajor axis, eccentricity and
oscillation in eccentricity ($\Delta e$) were the factors
examined. Jupiter's orbit must have had $|a - a_j| < 0.05$, $|e - e_j|
< 0.0156$, and $|\Delta e - \Delta e_j| < 0.0164$, while Saturn's
orbit required $|a - a_s| < 0.075$, $|e - e_s| < 0.0252$, and $\Delta
e - \Delta e_s| < 0.0256$.

We complemented our normal migration simulations with what we call
hereafter 'perfect migration' cases. In these simulations, the
planetesimals and embryos do not have any direct effect on the giant
planets, even during close encounters. However, their indirect effects
cannot be suppressed (specifically the H$_{\mathrm{sun}}$ term
  from eq. 32b. in Duncan et al. 1998) , but in principle they are
weaker. Thus the migration of the giant planets, starting with the
initial conditions from Brasser et al. 2009 (as in
Fig. \ref{migration}), met the above criteria in 3 out of 4
simulations. The giant planets had the full gravitational affect on
the planetesimals and embryos throughout these simulations, and the
mutual effects between planetesimals and embryos remained unchanged.

\section{Results}
We present the results of 12 simulations of terrestrial planet
formation each covering 150 Myr. Of these runs, 9 are normal migration
simulations and 3 'perfect migration' simulations (all simulations are
listed in Table 1. and refered to by run name, ``Test31'' etc.,
throughout). These two sets of simulations had qualitative and
quantitative similarities and are thus discussed at the same time and
combined in the figures.  First, the resulting planets are compared
with the current terrestrial planets, followed by a look at the
consequences the migration has on the structure of the asteroid belt.

\begin{figure}[h!]
\includegraphics*[width=8.4cm,angle=0]{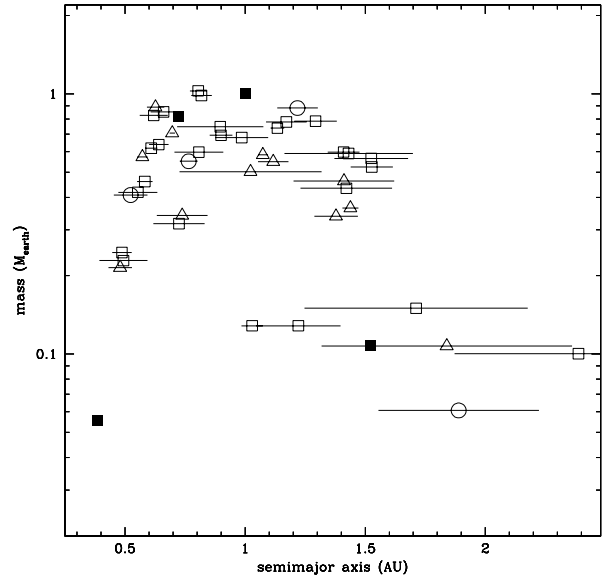}
\caption{The final mass ($M_\oplus$) for each planet produced in our
  simulations is plotted as a function of the planet's semimajor
  axis. The horizontal error bars show the locations of the perihelion
  and aphelion of the cooresponding orbit.  The open squares refer to
  the planets produced in the normal migration simulations, the open
  circles to the planets produced in the run with twice as many
  half-sized embryos, and the open triangles to those produced in the
  `perfect migration' simulations; the solid squares represent the
  real terrestrial planets.  }
\label{RandomMvS}
\end{figure}

\subsection{The planets}
Results for these simulations are summarized in Fig. \ref{RandomMvS},
where the final masses and semimajor axes of our synthetic planets are
compared to those of the real terrestrial planets (see also Table
\ref{runtable}). The trend is similar to that found in previous works
(see for instance Chambers et al. 2001), where the masses and
locations of Earth and Venus are nearly matched by a number of
different simulations, but most planets just exterior to Earth, near
$a$ $\sim$1.5 AU are at least 3 times more massive than Mars. However,
a handful of planets close to 1.8 AU were of similar mass to Mars. Of
note, Test31 had two $\sim$Mars-mass bodies, at 1.2 and 2.4 AU, with
an Earth mass planet at 1.52 AU. Test54, the only one of four
simulations starting with the smaller embryos with successful
migration, produced a sub-Mars mass body at 1.89 AU, just at the edge
of the current day asteroid belt. The Ran4 simulation produced a body
within 50\% of Mars' mass at 1.71 AU, though it had a high
eccentricity above 0.13 and was a member of a 3 planet system. In
general, planets produced at around 1.5 AU were $\sim$ 5 times more
massive than Mars, and Mars-mass bodies were typically only found
beyond 1.7 AU.

The total number of planets produced in each simulation is not
systematically consistent with the real terrestrial planet system.
Only two simulations produce 4 planets, where we define a ``planet''
as any embryo-sized or larger body with a semimajor axis less than 2.0
AU. Most simulations had 3 planets at the end, while one produced 5
planets. A common metric for measuring the distribution of mass among
multiple planets is the radial mass concentration statistic (RMC),
defined as

\begin{equation}
RMC = max\bigg(\frac{\sum M_j}{\sum M_j[\log_{10}(a/a_j)]^2}\bigg) ,
\end{equation}

\noindent where $M_j$ and $a_j$ are the mass and semimajor axis of
planet $j$ (Chambers, 2001). The bracketed function is calculated for
different $a$ in the region where the terrestrial planets form. The
RMC is infinite for a single planet system, and decreases as mass is
spread among multiple planets over a range of semimajor axes. The
current value of RMC for the solar system is 89.9. For all but one
simulation the RMC value is below the current solar systems value,
largely due to the large mass concentrated in a Mars-analog orbit 
  (we did not include the two embryos stranded in the asteroid belt
  region in these calculations, one in Test31 and one in
  TestPM24). The single simulation with a larger RMC value did not
have a Mars analog, and thus the mass was contained in a smaller
semimajor axis range.

The terrestrial planets have low eccentricities and inclinations,
Earth and Venus both have $e < 0.02$ and $i < 3^\circ$, properties which has
proved difficult to match in accretion simulations. O'Brien et
al. (2006) and Morishima et al. (2008) reproduced low eccentricities
and inclinations largely due to remaining planetesimals which damp the
orbital excitation of the planets.  A metric used as a diagnostic of
the degree of success of the simulations in reproducing the dynamical
excitation of the terrestrial planets is the angular momentum deficit
(AMD; Laskar 1997):

\begin{equation}
AMD = \frac{\sum_j M_j \sqrt{a_j}\left(1-\cos(i_j)\sqrt{1-e_j^2}\right)}{\sum_j M_j \sqrt{a_j}}  ,
\end{equation}

\noindent where $M_j$ and $a_j$ are again the mass and semimajor axis
and $i_j$ and $e_j$ are the inclination and eccentricity of planet
$j$. The AMD of the current solar system is 0.0014. The AMD for our
simulations ranged from 0.0011 to 0.0113. The plantesimal disk used in
these simulation is based on that from O'Brien (2006), and is
therefore not surprising that some AMDs are consistent with the solar
system value.  Simulation PM22 is the one with the largest final AMD,
because it produced an Earth-analog with a 10$^\circ$ inclination.

\begin{figure}
\includegraphics*[width=8.4cm,angle=0]{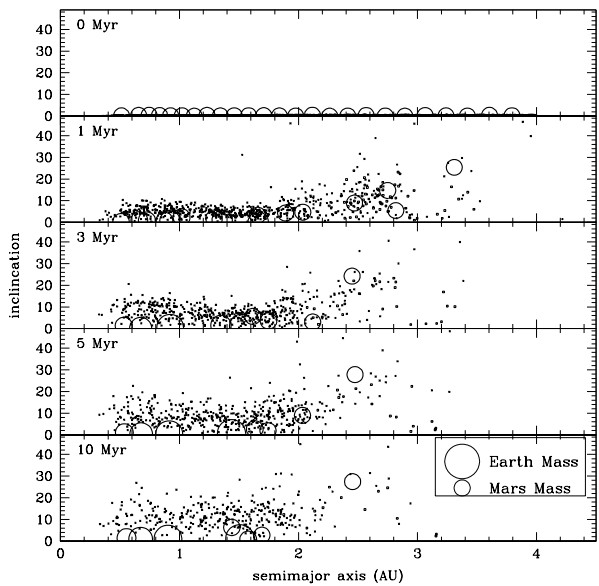}
\caption{Evolution of the system over time, showing the clearing of
  the asteroid belt region with inclination plotted as a function
  of semimajor axis. The open boxes are planetesimals on orbits within
  the current asteroid belt region, the crosses are planetesimals
  elsewhere, and the open circles are embryos or planets scaled in
  relation to their diameters. The simulation is Test31.  }
\label{31}
\end{figure}

\begin{figure}
\includegraphics*[width=8.4cm,angle=0]{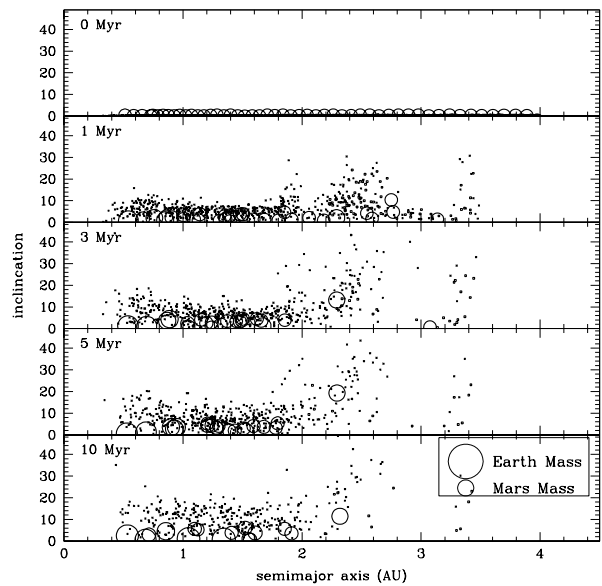}
\caption{Same as Fig.~\ref{31}, but for simulation Test54, which
  started from 50 embryos of 0.0467 $M_\oplus$
  instead of 25 embryos twice as massive.
}
\label{54}
\end{figure}

Figures \ref{31} and \ref{54} show snapshots of two systems evolving
over time. Of interest is the radial clearing caused by the movement
of the giant planets and the sweeping of their resonances,
particularly the $\nu_6$ resonance. This clearing progresses from the
outer edge of the disk towards the sun, following the migration of the
$\nu_6$ resonance, and stops at $\sim 2$~AU, which is the final
location of this resonance when the giant planets reach their current
orbits. Thus, the region of $a > 2.0$ is almost entirely cleared of
material in 10 Myr, with only handfuls of planetesimals surviving and
a single embryo. At 3 Myr, only $a > 2.5$ AU is largely cleared. 

\begin{figure}
\includegraphics*[width=8.4cm,angle=0]{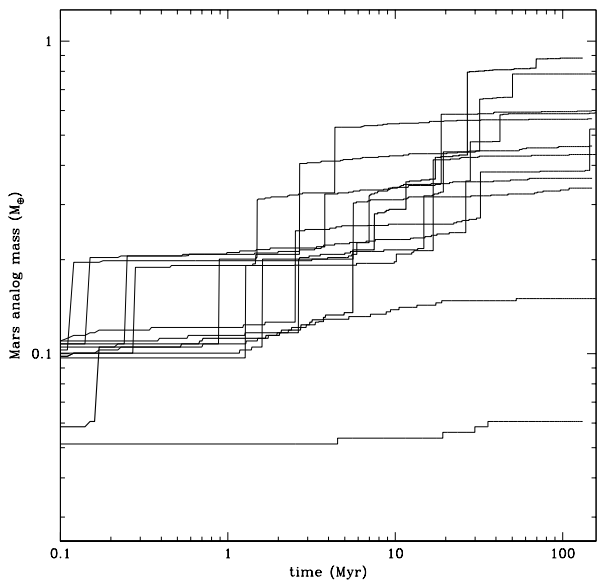}
\caption{Mass growth of the Mars analogs for all simulations plotted
  as a function of time. The most massive Mars analogs exceed the mass
  of Mars (0.11 $M_\oplus$) in only 2--3~Myr, and then in the next
  10--20 Myr continue to grow to their final sizes, ending many times
  more massive than Mars.  The two lines starting from $\sim 0.05$
  $M_\oplus$ are for two planets of simulation 'Test54', the only
  successful normal simulation that started with half-Mars mass
  embryos. The bold line shows the mass growth of the planet ending at
  $\sim 1.2$~AU; the thin line the planet ending at $\sim 1.9$~AU.  }
\label{marsgrowth}
\end{figure}

As seen in Figure \ref{marsgrowth} the accretion of embryos for the
Mars analogs (where a Mars analog is defined as the largest body
between 1.2--2.0 AU) begins immediately with $\sim$2 Mars-mass
typically being reached in only 2~Myr (note that Figure
\ref{marsgrowth} shows 12 growth curves, as there are two planets
displayed for Test54). Nine of the 11 Mars analogs have reached
0.2~$M_\oplus$ by 3 Myr. At 10~Myr 6 of the 11 have reached
0.3~$M_\oplus$, and by 30~Myr 10 of 11 are above 0.3~$M_\oplus$, or
$\sim$3~$M_ \mathrm{Mars}$. One might wonder if our inability to
produce a small Mars analog is due to the fact that, in all but one of
the presented simulations (Test54 is the exception), the planetary
embryos are initially $\sim$ one Mars mass. This is not regarded as a
problem for the following reasons.  First, the Mars analogs with
semimajor axes near that of Mars, near 1.5 AU, typically accreted 4 or
5 embryos; thus they consistently accreted much more mass than Mars,
and are not simply the result of a chance accretion between two Mars
mass embryos. Second, only two of the 11 Mars analogs did not accrete
another embryo, in Test54 and Ran4, but both had semimajor axes larger
than 1.7 AU, well beyond the current orbit of Mars. Third, our single
successful normal migration simulation that started with half-Mars
mass embryos also produced an Earth mass planet at 1.2~AU. This planet
was already two-mars masses in 5~Myr (notice that in the same
simulation one embryo escaped all collisions with other embryos and
therefore remained well below the mass of Mars - see Figure
\ref{marsgrowth}-- but this object ended up at 1.9~AU, well beyond the
real position of Mars). Finally, previous works (Chambers 2001;
Raymond et al. 2009; Morishima et al. 2010 to quote a few) which
started with embryos significantly less massive than Mars met the same
Mars-mass problem found here. The similarities between our work and
previous in terms of the mass distribution of the synthetic planets as
a function of semimajor axis suggest that the giant planet migration
does not affect significantly the terrestrial planet accretion
process. Thus it is unlikely that small changes in the adopted
evolution pattern of the giant planets could lead to significantly
different results.  Therefore the initial conditions do not appear to
be at fault for the failure to match the mass of Mars.

The reason for which the Mars analog consistently grows too massive is
twofold. First, they grow fast (in a few million years, as shown in
Figure \ref{marsgrowth}), compared with the timescale required to
effectively truncate the disk at $\sim 2$~AU (10~Myr, as shown in
Figures \ref{31} and \ref{54}). Second, the truncation of the disk
caused by the sweeping of the $\nu_6$ resonance is not sunward enough:
the final edge is approximately at 2~AU, whereas an edge at $\sim
1$~AU is needed (Hansen 2009; Kokubo et al. 2006; Chambers 2001).

\begin{figure}
\includegraphics*[width=8.4cm,angle=0]{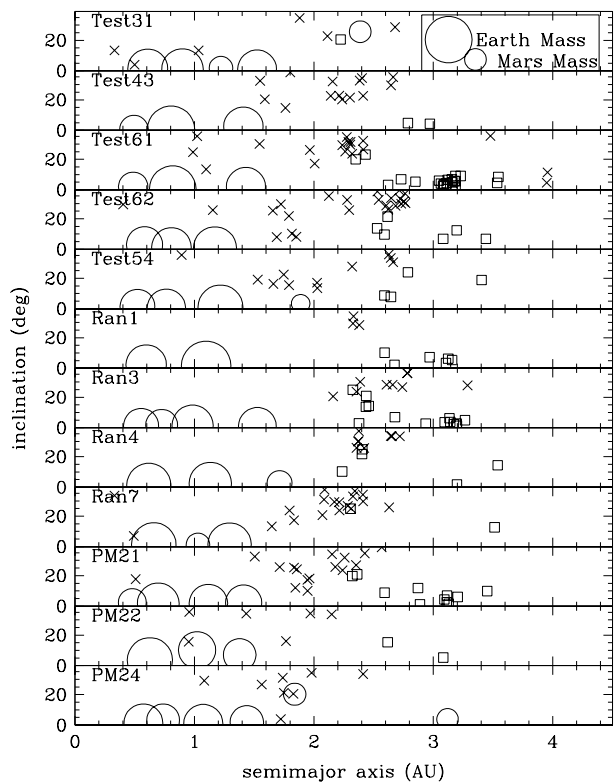}
\caption{Endstates of all simulations with the inclination plotted as
  a function of the semimajor axis with asteroids as open squares,
  non-asteroid planetesimals as crosses and embryos/planets as open
  circles scaled by their mass to the 1/3 power.
}
\label{allSims}
\end{figure}

Figure~\ref{allSims} shows the final incination
vs. semimajor axis distributions of all our simulations
(respectively, 'normal' and 'perfect' ones). The sizes of the symbols
representing the planets are proportional to the cubic roots of their
masses. Again, the problem of the mass of Mars stands out.

\subsection{The asteroid belt}

In the previous section we have shown that the an early sweeping of
secular resonances through the asteroid belt is not useful to solve
the small-Mars problem. Here we address the question of other
observational constraints. For this purpose, in this section we turn
to the asteroid belt, whose orbital distribution is very sensitive to
the effects of resonance sweeping (Gomes 1997; Nagasawa et al. 2000;
Minton \& Malhotra 2009; Morbidelli et al. 2010).

Morbidelli et al. (2010) have shown that the properties of the
asteroid belt after the slow migration of the giant planets are
largely incompatible with the current structure of the asteroid
belt. However, they assumed that the migration of the giant planets
occurred late, after the completion of the process of terrestrial
planet accretion and after the primordial depletion/dynamical
excitation of the asteroid belt. Thus, that work does not exclude the
possibility of an early migration. In fact, the outcome of an early
migration could be very different from that of a late migration for
two reasons: first, the initial orbits of the plantesimals are
quasi-circular and co-planar in the early migration case whereas they
are dynamically excited in the late migration case, which is an
important difference; second, planetary embryos reside in, or cross,
the asteroid belt region during the early time of terrestrial planet
formation, and this process has the potential of erasing some of the
currently unobserved signatures of resonace sweeping.

\begin{figure}
\includegraphics*[width=8.4cm,angle=0]{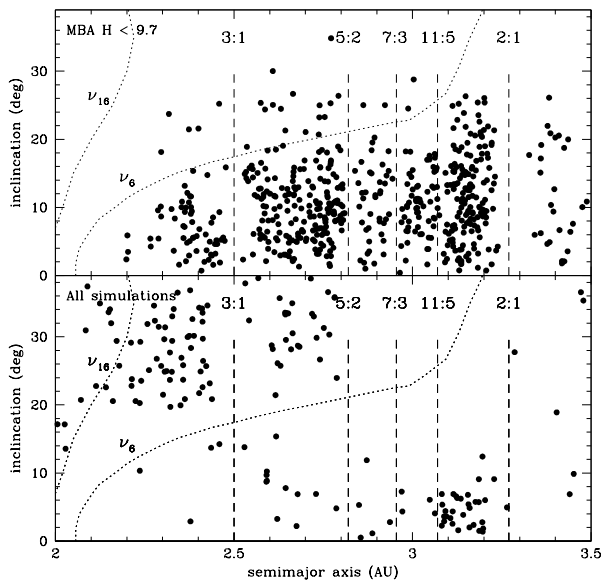}
\caption{(Top) The inclination of current day asteroids with absolute
  magnitude H $<$ 9.7, corresponding to D $\gtrsim$ 50 km, plotted as
  a function of their semimajor axis. The long-dashed lines show the
  location of the major mean motion resonances with Jupiter and the
  short-dashed curves the location of the $\nu_{6}$ and $\nu_{16}$ secular
  resonances. (Bottom) Surviving planetesimals from the 12 simulations,
  showing a strong depletion of low inclination bodies in the inner
  part of the asteroid belt region.}
\label{Asteroids}
\end{figure}

To compare the planetesimal distribution obtained in our simulations
with the ``real'' asteroid population, we focus on asteroids larger
than $\sim$50 km in diameter, as in previous works (Petit et al.
2001; Minton \& Malhotra 2009; Morbidelli et al. 2010). These
bodies are a reliable tracer of the structure of the asteroid belt
that resulted from the primordial sculpting process(es), as they are
too large to have their orbits altered significantly by the thermal
Yarkovsky effect or by collisions (see Fig. \ref{Asteroids},). Moreover, their
orbital distribution (see top panel of Fig~\ref{Asteroids}) is not
affected by observational biases, because all bodies of this size are
known (Jedicke et al. 2002).

The final distribution of the planetesimals residing in the asteroid
belt in our 12 simulations is shown in the bottom panel of
Fig~\ref{Asteroids}. As can be seen, the difference in orbital
distribution between the real belt and that resulting from the giant
planet migration process is striking. 

A simple metric used in Morbidelli et al. (2010) to quantify the
difference in orbital distributions between the real and the synthetic
belts is the ratio of asteroids above and below the location of the
$\nu_6$ secular resonance with semimajor axis below 2.8~AU. The current
day value for asteroids with a diameter above 50 km is 0.07. Combining
together all the surviving planetesimals from all our 12 simulations
results in a 67/13 ratio, in stark contrast to the current
value. Thus, our result is qualitatively similar to that of Morbidelli
et al. (2010), even though our resulting ratio is much larger than
that obtained in that work (close to 1/1).

The reason for the large ratio obtained in migration simulations, as
discussed in Morbidelli et al. (2010), is that the migration of the
giant planets forces the $\nu_6$ and $\nu_{16}$ secular resonances to
move Sun-ward.  More precisely, if the orbital separation of Jupiter
and Saturn increased by more than $1$~AU (as predicted by all models
and enacted in our simulations), the $\nu_6$ resonance sweeps the
entire asteroid belt as it moves inwards from 4.5~AU to 2~AU;
meanwhile the $\nu_{16}$ resonance sweeps the belt inside of 2.8~AU to
its current location at 1.9~AU (Gomes et al. 1997). In the inner
asteroid belt, the $\nu_{16}$ resonance sweeps first and the $\nu_6$
resonance sweeps second.  The $\nu_{16}$ resonance occurs when the
precession rate of the longitude of the node of the orbit of an
asteroid is equal to the precession rate of the node of the orbit of
Jupiter, and it affects the asteroid's orbital inclination. Given the
characteristic shape of the $\nu_6$ resonance location in the $(a,i)$
plane (see Fig~\ref{Asteroids}), the asteroids that acquire large
enough inclination when they are swept by the $\nu_{16}$ resonance,
avoid being swept by $\nu_6$; thus, their eccentricities are not
affected and they remain stable. Conversely, the asteroids that remain
at low-to-moderate inclinations after the $\nu_{16}$ sweeping are then
swept by the $\nu_6$ resonance and their eccentricities become large
enough to start crossing the terrestrial planet region. These bodies
are ultimately removed by the interaction with the (growing)
terrestrial planets. This process favors the survival of
high-inclination asteroids (above the current location of the $\nu_6$
resonance) over low-inclination asteroids and explains the large ratio
between these two populations obtained in the resonance sweeping
simulations. This ratio is larger in our simulations than in those of
Morbidelli et al. (2010), because the initial orbits of planetesimals
and embryos in our case have small inclinations and
eccentricities. Consequently, the secular resonance sweeping can only
increase eccentricities and inclinations. Conversely, in the
Morbidelli et al. (2010) simulations, the initial orbits covered a
wide range of eccentricities and inclinations.  Large eccentricities
or inclinations can be {\it decreased} by the secular resonance
sweeping. Thus, more objects could remain at low-to-moderate
inclinations after the $\nu_{16}$ sweeping and fewer objects were
removed by the $\nu_{6}$ sweeping than in our case.

We conclude from our simulations that the migration of the giant
planets with an $e$-folding time of 5~Myr (or longer, as the effects of
secular resonance sweeping increases with increasing migration timescale)
is inconsistent with the current structure of the asteroid belt, even
if it occurred early. In fact, our simulations provide evidence that
the planetary embryos crossing the asteroid belt
during the process of formation of the terrestrial planets are not
able to re-shuffle the asteroid orbital distribution and erase the
dramatic scars produced by secular resonance sweeping. 

\section{Discussion and Conclusions}

This paper has investigated the effects of {\it early} giant planet
migration on the inner disk of planetesimals and planetary embryos. In
the context of solar system formation, ``early'' is immediately
following the disappearence of the gas disk, which is identified as
time-zero in our simulations. The giant planets are migrated towards
their current orbits with a 5~Myr $e$-folding time which is
appropriate if the migration is caused by planetesimals scattering.

We have shown that the sweeping of secular resonances, driven by giant
planet migration, truncates the mass distribution of the inner disk,
providing it with an effective outer edge at about 2~AU after about
10~Myr.  This edge is too far from the Sun and forms too late to
assist in the formation of a small Mars analog. In fact, Chambers
(2001) already showed similar results starting from a disk of objects
with semi-major axes $0.3<a<2.0$~AU, the terrestrial planet accretion
process leads to the formation of planets that are systematically 3-5
times too massive at $\sim 1.5$~AU. For completeness, we have
continued our simulations well beyond the migration timescale of the
giant planets to follow the accretion of planets in the inner solar
system, and we have confirmed Chambers (2001) result.

Hansen (2009) showed that obtaining planets at $\sim 1.5$~AU that have
systematically one Mars mass requires that the disk of solid material
in the inner solar system had an outer edge at about 1~AU. The
inability of secular resonance sweeping during giant planets migration
to create such an edge suggests that a different mechanism needs to be
found.

Moreover, our study adds to the continuing inability of models with a
slow migration of the giant planets, $\tau \gtrsim 5$ Myr, to leave an
asteroid belt with a reasonable inclination distribution.  Morbidelli
et al. (2010) argued that the only possibility for the orbits of Jupiter
and Saturn to move away from each other on a timescale shorter than
1~Myr is that an ice giant planet (presumably Uranus or Neptune) is
first scattered inwards by Saturn and is subsequently scattered
outwards by Jupiter, so that the two giant planets recoil in opposite
directions. They dubbed this a ``jumping-Jupiter'' evolution and
showed that in this case the final orbital distribution of the
asteroid belt is consistent with that observed. Again, Morbidelli et
al. (2010) worked in the framework of a ``late'' displacement of the orbits
of the giant planets. Our results in this paper suggest that a
jumping-Jupiter evolution would also be needed in the framework of an
``early'' displacement of the orbits of the giant planets.

At this point, it is interesting to speculate what the effects of an
``early'' jumping-Jupiter evolution would be on the terrestrial
planet formation process. In essence, an early jumping-Jupiter
evolution would bring the giant planets to current orbits at a very
early time. So, the outcome of the terrestrial planet formation
process would resemble that of the simulations of Raymond et
al. (2009) with giant planets initially with their current orbital
configuration, labelled 'EJS' in that work. In these simulations,
though, (see their Fig.~10), the Mars analog is, again, systematically
too big. It is questionable whether a jumping-Jupiter evolution could
bring the giant planets onto orbits with current semimajor axes but
larger eccentricities, as required in the most successful simulations
of Raymond et al. (2009), labelled 'EEJS'. However, even though
jumping-Jupiter evolutions satisfying this requirement were found, it
is important to note that all of the outcomes of the EEJS simulations
of Raymond et al. (2009).  While producing a small Mars in several
cases, the EEJS simulations failed in general to bring enough water to
the terrestrial planets, formed the Earth too early compared to the
nominal timescale of 50~Myr and left the terrestrial planets on orbits
too dynamically excited.  For all these reasons an early
jumping-Jupiter evolution is not a promising venue to pursue for a
successful model of terrestrial planet formation.

In conclusion, our work substantiates the problem of the small mass of
Mars and suggests that understanding terrestrial planet formation
requires a paradigm shift in our view of the early evolution of the
solar system.

\subsection*{Acknowledgments}
The authors would like to thank an anonymous reviewer for a careful
reading of the manuscript.  KJW acknowledges both the Poincar\'{e}
Postdoctoral fellowship at the Observatoire de C\^{o}te d'Azur.  This
work is part of the Helmholtz Alliance's 'Planetary evolution and
Life', which KJW and AM thank for financial support. Computations were
carried out on the CRIMSON Beowulf cluster at OCA.

\section{Tables}

{ 
\small \include{Table} }

\section{References}

Agnor, C., Canup, R., Levison, H. 1999, Icarus 142, 219 \\
Batygin, K. \& Brown, M.~E. 2010, ApJ 716, 1323\\
Bouvier, A., Blichert-Toft, J., Moynier, F., Vervoort, J.~D., 
\& Albar{\`e}de, F.\ 2007, \gca 71, 1583 \\
Burkhardt, C., Kleine, T., Bourdon, B., Palme, H., Zipfel, J.,
Friedrich, J.~M., \& Ebel, D.~S.\ 2008, \gca 72, 6177 \\
Brasser, R., Morbidelli, A., Gomes, R., Tsiganis, K., \& Levison, H.~F. 2009, A\&A 507, 1053\\
Chambers, J. 2001, Icarus 152, 205 \\
Chambers, J. 2007, Icarus 189, 386 \\
Chambers, J.E. \& Wetherill G.W. 1998, Icarus 136, 304\\ 
Duncan, M.~J., Levison, H.~F. \& Lee, M.~H. 1998, ApJ 116, 2067\\
Fernandez, J.~A., and Ip, W. 1984, Icarus 58, 109 \\
Ferraz-Mello, S., Beaugé, C. \& Michtchenko, T.~A. 2003, CeMDA 87, 99\\
Gomes, R., Levison, H., Tsiganis, K. \& Morbidelli, A. 2005, Nature 435, 466\\
Gomes, R.~S. 1997, AJ 114, 396\\ 
Greenberg, R., Hartmann, W.K., Chapman, C.R. \& Wacker, J.F. 1978, Icarus 35, 1\\
Hansen B.~M.~S. 2009, ApJ 703, 1131\\
Jedicke, R., Larsen, J., \& Spahr, T.\ 2002. In: 'Asteroids III' (W.F. Bottke, A. Cellino, P. Paolicchi and R. P. Binzel, eds), Univ. Arizona Press, Tucson, Arizona. \\
Kenyon, S.J. \& Bromley, B.C. 2006, AJ 131, 1837\\
Kleine, T., Touboul, M. \& Bourdon, B. 2009, \gca\ 73, 5150 \\ 
Kley, W., Bitsch, B. \& Klahr, H. 2009, A\&A 506, 971\\
Kokubo, E. \& Genda, H. 2010, ApJ 714, L21\\
Kokubo, E., \& Ida, S.\ 1998, Icarus, 131, 171\\
Kokubo, E., \& Ida, S.\ 2000, Icarus, 143, 15\\
Kokubo, E., Kominami, J. \& Ida, S. 2006, ApJ 642, 1131\\
Kring, D.~A., \& Cohen, B.~A.\ 2002, JGRE, 107, 5009 \\
Laskar, J.\ 1997, \aap, 317, L75\\
Levison, H. F., Dones, L., Chapman, C. R., Stern, S. A., Duncan, M. J. \& Zahnle, K. 2001, Icarus 151, 286 \\
Levison, H.~F. \& Agnor, C. 2003, ApJ 125, 2692\\
Masset F. \& Snellgrove, M. 2001, MNRAS 320, 55\\
McNeil, D. Duncan, M. \& Levison, H.~F. 2005, ApJ 130, 2884 \\
Minton, D.~A., \& Malhotra, R.\ 2009, \nat, 457, 1109\\
Morbidelli A. \& Crida, A. 2007 Icarus 191, 158\\
Morbidelli, A., Brasser, R., Gomes, R., Levison, H.F. \& Tsiganis, K. 2010, AJ 140, 1391\\
Morishima, R., Schmidt, M.~W., Stadel, J., \& Moore, B.\ 2008, \apj 685, 1247\\
Morishima, R., Stadel, J. \& Moore, B. 2010, Icarus 207, 517\\
Nagasawa, M., Tanaka, H., \& Ida, S.\ 2000, \aj 119, 1480 \\
Nagasawa, M. \&  Lin, D.~N.~C. 2005, ApJ 632, 1140\\
O'Brien, P., Morbidelli, A. \& Levison, H. 2006, Icarus 184, 39\\
Peale, S.~J. \& Lee, M.~H. 2002, Science 298, 593\\
Petit, J.-M., Morbidelli, A. \& Chambers, J. 2001. Icarus 153, 338.\\
Pierens, A. \& Nelson, R. 2008, A\&A 482, 333\\
Raymond, S.~N., Quinn, T., \& Lunine, J.~I. 2004, Icarus 168, 1 \\
Raymond, S.~N., Quinn, T., \& Lunine, J.~I. 2005, ApJ 632, 670 \\
Raymond, S.~N., Quinn, T., \& Lunine, J.~I. 2006, Icarus 183, 265\\
Raymond, S.~N., Quinn, T., \& Lunine, J.~I. 2007, Astrobiology 7, 66 \\
Raymond,  S.~N.,O'Brien, D.~P., Morbidelli, A. \& Kaib, N.~A. 2009, Icarus 203, 644 \\
Ryder, G., Koeberl, C \& Mojzsis, S. 2000. In: 'Origin of the Earth and Moon' (R. Canup \& R. Knighter, eds). Univ. Arizona Press, Tucson, Arizona. \\
Ryder, G.\ 2002, Journal of Geophysical Research (Planets), 107, 5022 \\
Strom, R.~G., Malhotra, R., Ito, T., Fumi, Y. \& Kring, D.~A. 2005, Science 309, 1847\\
Tera, F., Papanastassiou, D.~A. \& Wasserburg, G.~J. 1974, E\&PSL 22, 1\\
Thommes, E.~W., Duncan, M.~J., \& Levison, H.~F.\ 1999, \nat\ 402, 635 \\
Thommes, E., Nagasawa, M. \& Lin, D.~N.~C. 2008, ApJ 676, 728\\
Thommes, E.~W., Bryden, G., Wu, Y., \& Rasio, F.~A.\ 2008b, \apj 675, 1538 \\
Tsiganis, K., Gomes, R., Morbidelli, A. \& Levison, H.  2005, Nature 435, 459\\
Wetherill, G.W. \& Stewart G.R. 1993, Icarus 106, 190\\

\end{document}

%% file: Table.tex
%\onecolumn
%\section*{Tables}
\renewcommand{\arraystretch}{1.0}

\begin{table}
\begin{center}
\begin{tabular}{l c c c c c c}
\hline
Run  & N & AMD & RMC & $M_\mathrm{planets}$& $M_\mathrm{Mars}$&$a_\mathrm{Mars}$\\
\hline
Normal Migration&   &      &      &      &     & \\ % 2/10 Total 9/30
Test31 & 5 &0.0085&32.69 &2.10  &0.56 &1.52\\
Test43  & 3 &0.0011&45.51 &1.85  &0.59 &1.41\\
Ran1 & 3 &0.0035 & 58.77&1.82& 0.43    &1.42  \\
Ran3 & 4 &0.0017 & 38.95&1.93& 0.52    &1.52 \\
Ran4 & 3 &0.0023 & 42.09&1.71& 0.14    &1.71 \\
Ran7 & 3 &0.0011 & 50.30&1.76& 0.78    &1.29 \\
\hline
Double Asteroids&    &      &      &      &     & \\ % 2/4
Test61  & 3 &0.0023&42.38 & 1.82 &0.59 &1.43\\ % Jup e ?
Test62  & 3 &0.0014&66.43 & 1.83 &  - & -  \\ %GREAT
\hline
Double Embryos&    &      &      &      &     & \\ %1/4
Test54 & 4 &0.0011& 40.53& 1.90 & 0.06 &1.88 \\ %638 or 502
\hline
Perfect Migration& &   &     &     & \\
PM21 & 4 & 0.0016 & 40.87 & 1.93& 0.46    &1.41 \\
PM22 &  3 & 0.0113 & 51.34 & 1.72& 0.33    &1.37 \\
PM24 &  6 & 0.0076 & 25.93 & 2.06& 0.36    &1.43 \\
\hline
MVEM & 4 & 0.0014 & 89.9  & 1.88 & 0.11 & 1.52 \\
\hline
\end{tabular}
\caption{Simulation results for each simulation included are the
  number of planets $N$, the angular momentum deficit (AMD) and the
  radial mass concentration (RMC) for each system of planets, the
  total mass of the planets $M_\mathrm{planets}$, and Mars analog
  $M_\mathrm{Mars}$ in Earth Masses, and the semimajor axis of the
  Mars analogs $a_\mathrm{Mars}$. The entry MVEM is data for the
  current terrestrial planets. Note that Test31 and PM24 each had one
  embryo stranded in the asteroid belt region, with $a > 2.0$ AU,
  which are counted in their $N$, but not used to calculate RMC or
  AMD.  }
% PM  % stuff
% from Stats.csh in
%PerfectMig/Plots/ on projekct
\label{runtable}
\end{center}
\end{table}